\documentclass[epj]{svjour}
\usepackage{amsmath,amssymb}

\usepackage{color}

\usepackage{graphicx}

\def \Cor {\mathfrak{C}}

\def \equals {\ = \ }

\newcommand{\onlinecite}[1]{\cite{#1}}

\begin{document}

\title{Physics behind the  minimum of  relative entropy measures for correlations}
\author{Karsten Held\inst{1} and Norbert J. Mauser \inst{2}}
\institute{Institute of Solid State Physics, Vienna University of Technology, 10
40 Vienna, Austria
\and Wolfgang Pauli Inst. c/o Fak. Mathematik, Univ. Wien, A-1090 Vienna}
\date{\today}

\abstract{
The relative entropy of a correlated state and an uncorrelated 
reference state is a reasonable measure for the degree of 
correlations. A key question is however which uncorrelated state 
to compare to. The relative entropy becomes
minimal for the uncorrelated reference state that has the 
same one-particle density matrix as the correlated state.
Hence, this particular measure, coined nonfreeness, is  unique and reasonable. 
We demonstrate that for relevant physical situations, such as 
finite temperatures
or a correlation enhanced orbital splitting, other choices of 
the uncorrelated state,
even educated guesses,
overestimate correlations.
}

%71.10.Fd  Lattice fermion models (Hubbard model, etc.)
%71.10.-w Theories and models of many-electron systems 
% 71.27.+a Strongly correlated electron systems; heavy fermions
\maketitle

%%%%%%%%%%%%%%%%%%%%%%%%%%%%%%%%%%%%%%%%%%%%%%%%%%%%%%%%%%%%%%%%%%%%%%%%%%%%%%%%%%%%%%%%%%%%%%%

%\section{Introduction}
 Correlated electrons give rise to fascinating physics such
as quantum criticality \cite{QC}, Mott-Hubbard transitions \cite{MHT},
or spin-fluctuations \cite{SF} e.g.\
in high-temperature superconductors. However, correlations are   
particularly difficult to deal with in theory and even a universally agreed definition, or measure, of "correlation"
is hitherto lacking. 
There is a general agreement that a Hartree-Fock Slater determinant represents an uncorrelated state, even though such a wave
function includes formally something one might call "correlations", which originate from the antisymmetrization
of the wave function.
Hence a correlation measure typically  considers the difference of
the correlated  state  vs.\  an uncorrelated  Hartree-Fock  calculation.
In this situation, the questions are: For what quantity should one consider the difference?
To which uncorrelated (possibly mixed) state should one compare to?

One possibility to quantify correlation is to look at the energy difference
\begin{equation}
E_{\rm corr}-E_{\rm free}
\end{equation}
between the (correlated) state investigated ($E_{\rm corr}$) and an uncorrelated (or free) state $E_{\rm free}$, which is also coined correlation energy. 
This is e.g.\ the typical quantity considered in quantum chemistry or 
density functional theory (DFT) \cite{Hohenberg64a,Jones89a,Martin04}.
%   where strictly speaking only the ground state  energy, its derivatives, and the density can be calculated. 
The exchange-correlation energy $E_{\rm xc}$ is, as the
difference to the Hartree energy, also readily accessible in DFT, at least 
within e.g.\ the local density approximation (LDA).  This requires,
however, still  a separation into exchange and correlation part.

In many-body theory on the other hand, one often considers two particle 
correlation functions of the type \cite{Abrikosov}
\begin{equation}
C_{ijkl}=\langle c^{\dagger}_i c_j^{\dagger}  c^{\phantom{\dagger}}_k c_l^{\phantom{\dagger}}    \rangle -  \big( \langle c^{\dagger}_i c_l^{\phantom{\dagger}}  \rangle\langle c^{\dagger}_j c_k^{\phantom{\dagger}}    \rangle - \langle c^{\dagger}_i c_k^{\phantom{\dagger}}  \rangle\langle c^{\dagger}_j c_l^{\phantom{\dagger}}    \rangle \big),\label{Eq:corr}
\end{equation}
where, $c^{\dagger}_i$  ($c^{\phantom{\dagger}}_i$) is the creation 
(annihilation) operator for a particle in the one-particle state $i$  (subsuming a possible 
spin index).
The second and third term subtract the uncorrelated expectation value. That is
a Slater determinant would yield as a result of the first term, the  latter two terms because of 
Wick's theorem \cite{Abrikosov} so that $C_{ijkl}=0$. Hence, we will
consider a state with  $C_{ijkl}\neq 0$ as ``correlated'' in the following.
The problem with Eq.\ (\ref{Eq:corr}) as a correlation measure is that there
is a myriad of such correlation functions - not only the two-particle correlation functions 
of Eq.\ (\ref{Eq:corr}) but also $n$-particle generalizations thereof. Even for a ``highly correlated'' state
{\em some} of the correlation functions can be small, even smaller than 
for what one would consider a "weakly  correlated" state.

%Note that here, and in the following, we consider equal-time (static) correlations.\\

Based on ideas from statistical or information theory,
 correlation measures have been proposed which are based 
on the 
$1$-particle density matrix (1PDM) or its eigenvalues. \cite{LichtnerGriffin,ZiescheEtAl,ZiescheEtAl2,Gori-GiorgiZiesche,GersdorfEtAl,Ziesche,EsquivelEtAl,GrobeRzazewskiEberly} 
%This class includes the ``nonidempotency" 
%of the 1PDM \cite{LichtnerGriffin,Ziesche}, the ``degree of correlation" 
%\cite{GrobeRzazewskiEberly}, and the ``information entropy" or ``correlation entropy" 
%\cite{Ziesche,EsquivelEtAl,GersdorfEtAl,ZiescheEtAl,ZiescheEtAl2,Gori-GiorgiZiesche}.
In Refs.\ \onlinecite{GottliebMauserPRL,Nonfreeness}, these concepts were extended to mixed states,
which requires the full density matrix instead of the 
1PDM. The resulting measure of correlation entropy, called ``nonfreeness,"  \cite{Nonfreeness}
is for a pure state equivalent to the
particle-hole symmetric correlation entropy introduced in Ref.\  \onlinecite{Gori-GiorgiZiesche}.
Beyond this, also mixed states such as thermal ensembles or open
subsystems can be treated. Such systems cannot be treated by a single Slater determinant but
behave  as a ``mixture" of pure states, best represented by the full density operator.

The basic idea behind  nonfreeness is that a many-body state which has the form of a grand canonical thermal equilibrium ensemble of  non-interacting particles has zero correlation.
Such a non-interacting state is called  ``free" in Ref.\  \onlinecite{Nonfreeness} and has  $C_{ijkl}=0$ in 
Eq.\ (\ref{Eq:corr}). Free states include all Slater determinant wavefunctions but also mixed states whose density operators satisfy a generalization of 
the (finite-temperature) Hatree-Fock ansatz.

For a many-electron state with density matrix $\rho$, there exists a unique free state $\Gamma_{\gamma_\rho}$ that has the same 1PDM $\gamma_{\rho}$ as  $\rho$. With this $\gamma_{\rho}$, Ref.\  \onlinecite{Nonfreeness} 
defines as  ``nonfreeness'' the relative entropy \cite{Vedral}
\begin{equation}
 \Cor(\rho) \equiv - {\rm Tr} \big(  \rho  \log\Gamma_{\gamma_\rho}  \big) +  {\rm Tr} \big(  \rho \log \rho \big)\label{Eq:nonfreeness} \;.
\end{equation}

Recently, the use of relative entropy
has also been proposed in Ref.\ \onlinecite{Byczuk} as a measure of correlations
. Here, correlations are quantified by the entropy relative to
physically motivated uncorrelated reference states with density matrix $\Gamma_{\rm ref}$ as \cite{Vedral}
\begin{equation}
S(\rho|\Gamma_{\rm ref}) \equiv - {\rm Tr} \big(\rho  \log  \Gamma_{\rm ref} \big) +  {\rm Tr} \big(  \rho \log \rho \big).
\label{Eq:Refstate}
\end{equation} 
 This measure has been applied\cite{Byczuk}  to the Hubbard model, MnO, FeO, CoO and NiO.
As examples for the  reference states $\Gamma_{\rm ref}$, the paramagnetic, antiferromagnetic or LDA
Slater determinants \cite{Byczuk} were taken
(restricted to a single site subsystem). 
Obviously, there is an ambiguity which
 physically motivated reference state to choose.
Hence, Ref.\  \onlinecite{Byczuk} proposes to
minimize the relative entropy, which was however considered
a too hard computational problem.

%$S(\Delta|\Gamma_{\gamma_\Delta})$ the ``nonfreeness of $\Delta$" and denoted it by 
%\begin{equation}
% \Cor(\Delta) \equals S(\Delta|\Gamma_{\gamma_\Delta}).
%\end{equation}
%As shown in Proposition~\ref{how to compute nonfreeness} below, it is as easy (or difficult) to compute $\Cor(\Delta)$  as it is to diagonalize $\Delta$.  
%The nonfreeness $\Cor(\Delta)$ provides an information theoretic measure of the  ``correlation in" $\Delta$, in the sense of its deviation from a state of independent-particle form.  
%As shown in \cite{Nonfreeness}, the nonfreeness of the state of a many-electron system is greater than or equal to that of any of its subsystems.  

 Obviously,  $\Cor(\rho)$ is the relative entropy for a particular 
reference state, i.e., $\Cor(\rho)= S(\rho|\Gamma_{\gamma_\rho})$.
It has been proven recently\cite{GottliebPC} that the nonfreeness is indeed the minimum
over all free  reference states $\Gamma_{\rm ref}$, which are uncorrelated in the sense that  $C_{ijkl}=0$ in Eq.\ (\ref{Eq:corr}):
\begin{equation} 
  \Cor(\rho)\equiv S(\rho|\Gamma_{\gamma_\rho}) \equals \min_{\Gamma_{\rm ref}} S(\rho|\Gamma_{\rm ref}).
\label{Eq:min}
\end{equation}
%Thus the nonfreeness provides a lower bound to the $S(\rho|\Gamma_{\rm ref})$ correlation measures. 

In this paper, we argue that this minimal relative entropy is also most reasonable from a physical
point of view: Its reference state $\Gamma_{\gamma_\rho}$ has the same 1PDM, i.e., all one-particle expectation values
  $\langle c^{\dagger}_i c_j \rangle$ are the same.  From this 1PDM and   Wick's theorem
for an uncorrelated state with $C_{ijkl}=0$,
all correlation functions and hence the full density matrix $\Gamma_{\gamma_\rho}$can be calculated, see Appendix.
Let us also  emphasize, that nonfreeness $\Cor(\rho)$ defines this way a {\em unique} relative entropy measure of correlations.
Obviously, it measures correlations of a state with given 
density matrix $\rho$, but not how correlated a  Hamiltonian is. This
is reasonable if we want to quantify the correlations of a (mixed) state,
independently on whether this is e.g.\
 the groundstate of one Hamiltonian
or the non-equilibrium state of another Hamiltonian.

%Nonfreeness is essentially a ``correlation entropy" measure that is especially suitable for open subsystems of many-electron systems.  In particular, nonfreeness lends itself to the kinds of applications considered in \cite{Byczuk}, where the open subsystems consisting of sites or blocks of sites in fermion lattices.  As an application of nonfreeness to a realistic lattice fermion systems, we will present a DMFT+LDA model of a transition metal oxide, compute the nonfreeness of the state at a single lattice site.  We will argue that the resulting measure is more sensible than ones obtained by using (\ref{Eq:min}) even with uncorrelated reference states $\Gamma_{\rm ref}$ that --  {\it prima facie} -- appear quite reasonable.  

% The entanglement of sites or blocks of sites within a fermion lattice has been the subject of much interesting study 

%The rest of this article is organized as follows.   In Section \ref{Sec:application}, we will apply nonfreeness to LaNiO$_3$/LaAlO$_3$ heterostructures, illustrating the difference of 
%this relative entopy measure to other possible choices. Finally, we summarize
%the main results in
% Section \ref{Sec:conclusion}.

%%%%%%%%%%%%%%%%%%%%%%%%%%%%%%%%%%%%%%%

%\section{Physics behind appropriate free state}
%\label{physics section}

Let us now discuss two physical examples, where
even a well educated choice of the uncorrelated reference
state leads to an overestimation of ``correlation'' compared
to the corresponding non-free state.

{\bf (i) Finite temperatures.}
%\subsection{Finite temperatures}
Let us assume we have, for example, an antiferromagnet. Here one might 
be inclined to take an antiferromagnetic Slater determinant such as 
the ground state of a Hartree-Fock calculation for the same Hamiltonian
as reference state.
However, at finite temperatures the antiferromagnetic magnetization is reduced.
Hence, if we employ the relative entropy Eq.\ (\ref{Eq:Refstate}) between
 this finite-temperature Hartree-Fock solution
and  the  antiferromagnetic zero-temperature Slater determinant, it 
is finite. That is, Eq.\ (\ref{Eq:Refstate})  indicates the presence of correlations in a situation 
were there are none, in the sense that all $C_{ijkl}=0$.
In contrast, the  nonfreeness measure employs in this situation
the very same  finite-temperature Hartree-Fock state as a reference state
and  is hence zero.

Even if we consider a single Slater determinant with the correct
(here finite-temperature Hartree-Fock) magnetization, its density matrix
is different from that of the finite temperature Hartree-Fock 
calculations which represents a thermal ensemble of Slater determinants.
Hence, according to Eq. (\ref{Eq:Refstate}) we would, even with a more well
educated guess for the reference state,  call the finite-temperature 
Hartree-Fock state correlated.

If truly correlated states are considered, the situation becomes more complicated,
but the nonfreeness warranties that we compare to a  reference state which has
the correct, finite temperature  magnetization and which is an uncorrelated
ensemble being more general than a single Slater determinant. This free reference state is 
as close to the correlated state as possible: it has the same 1PDM and
from requiring $C_{ijkl}=0$ in  Eq.\ (\ref{Eq:Refstate}) and its $N$-particle generalizations 
follows the full density matrix.

%\subsection{Orbital level splitting}
{\bf (ii) Orbital splitting.}
Let us as a second example consider a typical solid state situation.
There are orbitals which are split
by a one-particle  crystal field.
This crystal field as well as the hopping between different sites 
can be calculated for example by LDA.

As a specific example, we will take a LaNiO$_3$/LaAlO$_3$  heterostructure
for which the  low-energy  non-interacting  LDA Hamiltonian is explicitly given in Ref.\ \onlinecite{Hansmann}.
Since Ni has an open $d$-shell with, on average, one electron in two $e_g$ orbitals,
there is a strong Coulomb interaction for the case that there is a second
electron in the  $e_g$ orbitals on the same site. As a consequence,
strong electronic correlations can emerge. These have been calculated 
in Refs.\ \onlinecite{HansmannPRL,Hansmann} employing the LDA +  dynamical mean field theory (DMFT) approach.
\cite{LDADMFT1,LDADMFT2,LDADMFT3,LDADMFT4}

In  LDA there are somewhat 
more $x^2-y^2$  than  $3z^2-r^2$  electrons. We express this by the one-particle occupations: $n_1=0.376$ for each spin compared to $n_2=0.124$,\cite{Hansmann} where 1 and 2 denote the  $x^2-y^2$  and $3z^2-r^2$ orbital, respectively, and $n_1,n_2$ is the same for spin-up and -down in the paramagnetic phase considered. This results in the 1PDM 
\begin{equation}
 \gamma=\left( 
\begin{array}{c c c c}
n_1 &  & & \\
  & n_1 & & \\
  &  &n_2 & \\
  &  &  &n_2 
\end{array}
\right) \;\;\;\; ,
\label{Eq:1PDM}
\end{equation}
which is  diagonal
for the given Hamiltonian.

If the Coulomb interaction $U$ (and Hund's exchange $J$) is taken into account, the effective crystal field splitting between the two orbitals is 
enhanced \cite{Hansmann}. While the increased orbital splitting
is an effect of $U$ it does  not necessarily imply a correlated state.
In fact, such physics can already be described in an uncorrelated
Hartree-Fock (or 
LDA+U \cite{LDAU}) calculation, where the splitting is enhanced 
by $U (n_1-n_2)$ . For the LDA+U solution, we have a free state with zero nonfreeness,
whereas the relative entropy to the $U=0$ Slater determinant as reference state would be finite.

Also for a truly correlated state, obtained  e.g.\  from a LDA+DMFT calculation, 
we will get an enhanced  effective crystal field splitting and different orbital occupations:
for  $U=5.8\,$eV  and inverse temperature $\beta=25\,$eV$^{-1}$ the orbital occupation
changes dramatically to
  $n_{1}=0.48835$ and $n_{2}=0.01200$.\cite{Hansmann}
Such a change could be the result of a completely uncorrelated, e.g. LDA+U, 
wave function or due to true electronic correlations.
Equal-time two-particle correlations can be described by 
pair-wise double occupations $d_{ij}=\langle c^{\dagger}_i c_i^{\phantom{\dagger}}  c^{\dagger}_j c_j^{\phantom{\dagger}}\rangle$ with the index $i$ subsuming the orbital and spin, more specifically the correlation is given by the difference $d_{ij}-n_in_j$, see Eq.\ (\ref{Eq:corr}).

In  Ref.\ \onlinecite{Hansmann}
the pairwise double occupations  have been calculated. For  $U=5.8\,$eV  and inverse temperature $\beta=25\,$eV$^{-1}$ one has e.g.:
$d_{1\uparrow 1 \downarrow}=0.01014$, $d_{2\uparrow 2 \downarrow}= 0.00003$, $d_{1\uparrow 2 \downarrow}= d_{2\uparrow 1 \downarrow}=0.00317$,  $d_{1\uparrow 2 \uparrow}= d_{1\downarrow 2 \downarrow}=0.00452$. 
If we neglect occupations with three or four electrons on a site (which are extremely rare for larger $U$),
we obtain from the pairwise double occupations and orbital occupations  
the   local  density matrix for a single site:
\begin{eqnarray*}
 \rho =\left( 
\begin{array}{c c c c c c c c c c c}
   \!\! n_\emptyset \!   & & & & & & & & & & \\
  & \!  \! \!  \tilde{n}_1 \!  & & & & & & & & & \\
  & & \!  \!  \! \tilde{n}_1 \!  & & & & & & & & \\
  & & & \!  \! \!  \tilde{n}_2 \!  & & & & & & &  \\
  & & & & \!  \! \!  \tilde{n}_2 \!  & & & & & & \\
  & & & & & \!  \! \!    \!  d_{1\uparrow 1 \downarrow}  \!  \! \!  \! & & & & & \\
  & & & & & &  \!  \!\!   \! d_{1\uparrow 2 \uparrow}   \!  \! \!  \! & & & & \\
  & & & & & & & \!  \! \!   \! d_{1\uparrow 2 \downarrow}  \!  \! \!  \! & & & \\
  & & & & & & & & \!  \! \!   \!  d_{1\downarrow 2 \uparrow}   \!  \! \!  \! & & \\
  & & & & & & & & & \!  \! \!   \!  d_{1\downarrow 2 \downarrow}   \!  \! \!  \! &  \\
  & & & & & & & & & & \!  \! \!    \! d_{2\downarrow 2 \uparrow}     \! \! \!  
\end{array}
\right) &   \!& 
\begin{array}{l}
\leftarrow \; | 0000 \rangle\\
\leftarrow \; | 1000 \rangle\\
\leftarrow \; |  0100 \rangle\\
\leftarrow \; |  0010 \rangle\\
\leftarrow \; |  0001 \rangle\\
\leftarrow \; | 1100 \rangle\\
\leftarrow \; | 1010 \rangle\\
\leftarrow \; | 1001 \rangle\\
\leftarrow \; |  0110 \rangle\\
\leftarrow \; |  0101 \rangle\\
\leftarrow \; |  0011 \rangle\\
\end{array}
\label{Eq:DMcor}\\
&   \!&
\begin{array}{l}
|1\!\!\uparrow\! 1\!\!\downarrow\! 2\!\!\uparrow\! 2\!\!\downarrow  \rangle
\end{array}
\end{eqnarray*}
Here, the right hand side indicates the meaning of the individual rows (and columns) of the density matrix, using an occupation number formalism with states as indicated in the last line.
Note that the symmetry of the Hamiltonian yields a purely diagonal density matrix
in our case.
For the one-particle sector of the density matrix, we need to subtract from the average occupation number all double occupations which involve the given orbital and spin
for obtaining the single particle occupation of this state,
i.e.,
 $\tilde{n}_1=n_1\! - \! d_{1\uparrow 1 \downarrow}\! - \! d_{1\uparrow 2 \uparrow}  d_{1\uparrow 2 \downarrow}$
and equivalently ($1 \leftrightarrow 2$) for $\tilde{n}_2$.
The zero occupation sector is given by the sum rule (neglecting occupations with three or four electrons):
 $n_\emptyset = 1\! - \!2\tilde{n}_1\! - \!2\tilde{n}_2\! - \!d_{1\uparrow 1 \downarrow}\! - \!d_{2\uparrow 2 \downarrow}
\! - \! d_{1\uparrow 2 \downarrow}\! - \! d_{1\uparrow 2 \uparrow}\! - \! d_{1\downarrow 2 \uparrow}\! - \! d_{1\downarrow 2 \downarrow}$.

For this local density matrix, we have calculated as outlined in the Appendix the relative entropy Eq.\ (\ref{Eq:Refstate})
vs.\ (i) the uncorrelated $U=0$ state which is defined by its one particle density matrix Eq.\   (\ref{Eq:1PDM}) 
and (ii) the free state with the same 1PDM as the LDA+DMFT correlated state. The latter relative entropy is the nonfreeness.

\begin{figure}[t]
\includegraphics[width=8.1cm]{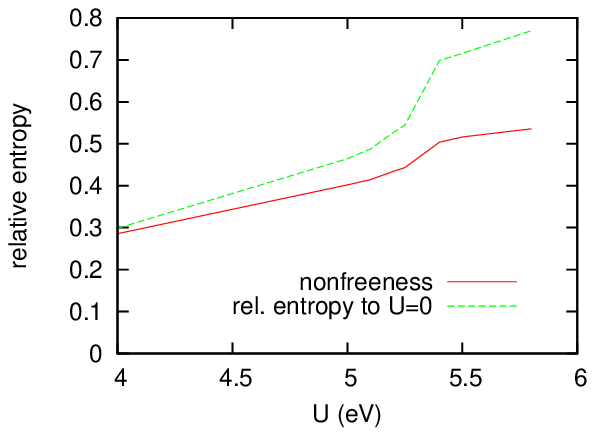}
\caption{(Color online) Relative entropy  vs.\ Coulomb interaction $U$ 
of the correlated DMFT state of  LaNiO$_3$/LaAlO$_3$  against (i) (green dashed) the uncorrelated $U=0$ state and (ii)  (red solid) the 
uncorrelated state with the same 1PDM (nonfreeness).
\label{Fig:entropy}
}
\end{figure}

Fig.\ \ref{Fig:entropy} shows the obtained relative entropy in the vicinity of the Mott-Hubbard metal insulator transition. In this parameter regime, double occupations are further suppressed so that electronic correlations are enhanced. This can be seen in the nonfreeness measure.
The relative entropy against the $U=0$ state increases however even more strongly.
This additional increase is not caused by a correlated wave function or
density matrix but by a shift in the occupation of the orbitals which can be described
in an uncorrelated state as well.

%\section{Conclusion}
{\bf Conclusion and Outlook.}
%\label{Sec:conclusion}
Recently, there has been a growing interest in the density matrix based measures of electronic correlations \cite{Nonfreeness,Byczuk,Horsch,Thunstrom}.
To this end, the entropy of the correlated state, which is described by its density matrix, has been compared to different reference states. The relative entropy is minimal for a particular uncorrelated (``free'') state: the one that has the same 1PDM and, hence, the same occupations of the (natural) orbitals. This measure is unique and coined  ``nonfreeness''.
If one compares to other uncorrelated reference states one gets a higher entropy measure of correlation. This difference simply stems from a different orbital occupation which does not necessitate true correlations, i.e., nonzero correlation functions  $C_{ijkl}\neq 0$.

As illustrating example we have considered finite temperatures and a two-orbital LaNiO$_3$/LaMnO$_3$ heterostructure. In this material, the Coulomb interaction $U$ changes (among others) the orbital occupations, effectively enhancing the crystal field splitting between the two orbitals. However, such a change of the orbital occupations does not require a correlated state. Indeed also a Hartree-Fock or LDA+U state shows this kind of physics. For this 
finite-$U$ Hartree-Fock state all correlation functions, such as those of Eq.\ (\ref{Eq:corr}), are zero, while the relative entropy with the $U=0$ state as a 
reference state would suggest correlations.

%Physically motivated choices for the uncorrelated references state might be the $U=0$ or finite-$U$ Hartree-Fock reference state. However, in comparison to the nonfreeness reference state, i.e., the one with the same (natural) orbital occupations (the same 1PDM) one gets a higher relative entropy, suggesting a higher degree of electronic correlations. This excess of relative entropy 
% simply stems from changes of the (natural) 
%orbital occupations which can occur inj an uncorrealted state as well.

Hence, we feel that the nonfreeness is the most suitable relative entropy measure for
electronic correlations if these are defined as
 corresponding to 
  nonzero correlation functions  $C_{ijkl}$ in
  Eq.\ (\ref{Eq:corr}), which is actually a usual definition for correlations in solid state theory: Any state with $C_{ijkl}=0$ and $N$-particle generalizations thereof has  zero nonfreeness.
Nonfreeness does however not distinguish  between genuine quantum correlations
and classical ones, which have been individuated  in Ref.\ \onlinecite{Thunstrom}.
For example in the paramagnetic phase, the single-site-reduced density matrix $\rho$ of the one-band Hubbard model
for half-filling and  double occupation $d$ and the free state $\Gamma_{\gamma_\rho}$ with the same 1PDM  are (in the single-site spin-up/-down basis as indicated on the right hand side)
\begin{equation}
\!\rho\!=\! 
\left(
\begin{array}{cccc}
d &&&\\
&\!\! \frac{1\!-\!2d}{2}\! \! &&\\
&& \!\! \frac{1\!-\!2d}{2} \!\! &\\
&&&d\\
\end{array} \right)\! , \;
\Gamma_{\gamma_\rho}\!= \! 
\left(
\begin{array}{cccc}
\!\frac{1}{4}\! &&&\\
&\!\frac{1}{4}\!  &&\\
&&\! \frac{1}{4}\! &\\
&&&\!\frac{1}{4}\!\\
\end{array} \right) \; \;
\begin{array}{l}
\leftarrow \; | 00 \rangle\\
\leftarrow \; | 10 \rangle\\
\leftarrow \; |  01 \rangle\\
\leftarrow \; |  11 \rangle\\
\end{array} .
  \end{equation}
For $U\rightarrow \infty$, we have $d\rightarrow 0$ and so that the nonfreeness becomes  $\Cor(\rho)=-\log(1/4)+\log(1/2)$. However, there are no genuine quantum correlations in the atomic limit. We can describe this state as a classical ensemble with 50\% probability of a spin-up and -down occupation.

{\bf Acknowledgments.}
We thank A. Gottlieb, P. Hansmann, P. Thunstrom, F. Verstraete, and D. Vollhardt for very helpful discussions and the Austrian Science Fund (FWF) through SFB ViCom F41 for financial support.
% K.H. also received support from  the
%FWF via Research Unit FOR-1346 of the German Science Fou(FWF I597), and N.M. from the FWF-ANR project I830.

{\bf Appendix.}
Let us add here some useful relations for the practical calculation
of relative entropies.
Let us consider the L\"owdin natural orbital\cite{Lowdin} basis where the 
1PDM is diagonal and the entries of the 1PDM are simply the probabilities 
$p_i =\langle c^{\dagger}_i c_i^{\phantom{\dagger}} \rangle$ for the occupation of orbital $i$. 
 For an uncorrelated or free state $\Gamma_{\rm free}$ these
$p_i$'s determine, because of  $C_{ijkl}=0$ in Eq.\ (\ref{Eq:Refstate}),
all correlated expectation values such as $\langle c^{\dagger}_i c_j^{\dagger}  c^{\phantom{\dagger}}_k c_l^{\phantom{\dagger}}    \rangle$ and hence also the full
density matrix  $\Gamma_{\rm free}$.
In a many-particle occupation-number basis
$|n_1 \cdots n_n \rangle$ with $n_i\in\{0,1\}$ denoting the occupation
of the  
 L\"owdin natural orbital $i$, the free density matrix $\Gamma_{\rm free}$ is also diagonal  with diagonal elements \cite{Nonfreeness}
$\prod_{i=1}^n p_i^{n_i} (1-p_i)^{1-n_i}$.
That is, we simply have the product of the probabilities to occupy or not to occupy orbital $i$.

For such a  $\Gamma_{\rm free}$, $-{\rm Tr} \{ \rho\log \Gamma_{\rm free}\}$ 
can be calculated  and shown to be minimal\cite{GottliebPC} for 
$\Gamma_{\rm free}=\Gamma_{\gamma_{\rho}}$, i.e., the uncorrelated free 
state with the same 1PDM as $\rho$. For this minimum, one obtains
\cite{Nonfreeness}
\begin{equation}
-{\rm Tr} \{ \rho\log \Gamma_{\gamma_{\rho}} \} = -\sum_i p_i \log (p_i) -\sum_i (1-p_i) \log  (1-p_i)
\end{equation}
so that the nonfreeness  $\Cor(\rho)$ in Eq.\ (\ref{Eq:nonfreeness})
can be easily calculated for a given density matrix $\rho$.

\end{document}